# NONLINEAR ACCELERATOR LATTICES WITH ONE AND TWO ANALYTIC INVARIANTS

V. Danilov

*Spallation Neutron Source Project, Oak Ridge National Laboratory,*

*Oak Ridge, TN 37830*

S. Nagaitsev

*Fermi National Accelerator Laboratory, Batavia, IL 60510*

## Abstract

Integrable systems appeared in physics long ago at the onset of classical dynamics with examples being Kepler's and other famous problems. Unfortunately, the majority of nonlinear problems turned out to be nonintegrable. In accelerator terms, any 2D nonlinear nonintegrable mapping produces chaotic motion and a complex network of stable and unstable resonances. Nevertheless, in the proximity of an integrable system the full volume of such a chaotic network is small. Thus, the integrable nonlinear motion in accelerators has the potential to introduce a large betatron tune spread to suppress instabilities and to mitigate the effects of space charge and magnetic field errors. To create such an accelerator lattice one has to find magnetic and electrtic field combinations leading to a stable integrable motion. This paper presents families of lattices with one invariant where bounded motion can be easily created in large volumes of the phase space. In addition, it presents 3 families of integrable nonlinear accelerator lattices, realizable with longitudinal-coordinate-dependent magnetic or electric fields with the stable nonlinear motion, which can be solved in terms of separable variables.

## I. Introduction

All present accelerators (and storage rings) are built to have "linear" focusing optics (also called lattice). The lattice design incorporates dipole magnets to bend particle trajectory and quadrupoles to keep particles stable around the reference orbit. These are "linear" elements because the transverse force is proportional to the particle displacement, $x$ and $y$. Only a fraction of the accelerator circumference is occupied by dipoles and quadrupoles, thus the equations of motion (in the uncoupled case) are written as:

$$\begin{cases} x'' + K_x(s)x = 0 \\ y'' + K_y(s)y = 0 \end{cases} \quad (1)$$
$$K_{x,y}(s+C) = K_{x,y}(s)$$

where $K_x$ and $K_y$ are piecewise constant functions of $s$ (the time-equivalent longitudinal coordinate), and $C$ is the accelerator circumference and the longitudinal motion is negligible.

One can notice that these are equations of two time-dependent uncoupled harmonic oscillators. Such an equation was first solved by Ermakov [1] who obtained its invariant, which in accelerator physics is called the Courant-Snyder invariant [2]. This invariant is most easily understood by introducing the so-called normalized phase-space coordinates:

$$z_N = \frac{z}{\sqrt{\beta(s)}},$$
$$p_N = p\sqrt{\beta(s)} - \frac{\beta'(s)z}{2\sqrt{\beta(s)}}, \qquad (2)$$

where $z$ stands for either $x$ or $y$, $p$ is similarly either $p_x$ or $p_y$, and $\beta(s)$ is either the horizontal or vertical beta-function (defined in Eq. (7)). In these new normalized variables, the initial time-dependent Hamiltonian associated with Eqs. (1) becomes time-independent,

$$H = \frac{1}{2}\left(p_N^2 + z_N^2\right), \qquad (3)$$

and thus leads to two invariants, the horizontal and vertical Hamiltonians. According to Eq. (3), in a linear lattice, all particles execute harmonic oscillations around the reference orbit with a frequency, known as the betatron tune, which is identical for all particles, regardless of their amplitude. Linear lattices have been considered attractive, in part because linear dynamics is easily understood.

However, several things make the perfect linear lattice undesirable. First, linear motion is unstable to perturbations in the focusing fields because of linear and non-linear resonances. Second, the lattice focusing strength depends on particle's kinetic energy deviation from the design beam energy. This effect is called chromaticity (or tune dependence on beam energy). In a perfectly linear lattice this undesirable effect could be quite large. The chromaticity is routinely corrected by non-linear focusing magnets, called sextupoles (the force proportional to $z^2$). Finally, ring designers often add higher multi-pole nonlinear elements to lattices. For example, a focusing element called an octupole ($\sim z^3$) is often added to increase the betatron tune dependence on the particle amplitude in order to achieve the so-called Landau damping by reducing the number of resonant particles. All these nonlinear (polynomial in $x$ and $y$) elements have one thing in common: they limit the available phase space area where the particle motion is regular (non-chaotic) because they are non-integrable.

In this paper we present several non-linear integrable accelerator lattices with regular particle orbits in a large phase-space volume around the reference orbit. The goal is to find such a combination of static electro-magnetic fields along the orbit that leads to two functionally independent invariants which are in involution (i.e. with zero Poisson brackets). Ultimately, the idea is to construct an accelerator with a large betatron frequency spread (> 10%) while maintaining regular (non-chaotic) particle orbits.

An overview of the advances in accelerator nonlinear lattices with integrable motion can be found in Ref. [3]. Ref. [3] outlined the main difficulty in finding such lattices: the fields must obey the stationary Maxwell equations (or the Laplace equation) and this is too severe of a constraint for already extremely rare cases of 2D integrable time-dependent classical Hamiltonians that describe the transverse motion of particles in accelerators. Even when the exact cases were found, the motion was unstable. Let us take, for example, time independent magnetic fields near a reference particle closed orbit. For static constant (along the orbit) fields, the time-independent Hamiltonian is automatically the system's invariant, but such a system would lack the 2D stability near the closed orbit because electric or magnetic fields cannot focus particles simultaneously in both directions (from here on we exclude weak focusing based solely on bending magnets). Only time-dependent strong focusing can produce stable linear motion. A nearly integrable nonlinear lattice with thin lenses was suggested in [3] where the motion is separable in Cartesian coordinates.

In this paper we adopt a different approach – the lenses are taken to be not thin. The special time (or longitudinal coordinate) dependence of the transverse fields can be chosen such that the 2D motion has one or two invariants in involution but the motion is essentially coupled as compared to the case in Ref. [3]. In the case of two invariants the motion is separable in spherical, parabolic, and elliptical coordinates and has large 4d volumes of stable nonlinear motion. In the next section we present the acceptable field equations. Then we introduce the special time-dependence of the fields to guarantee the existence of the motion invariants. The analysis of exactly integrable cases is done at the end of the paper.

## II. Propeties of 2D Fields

As we mentioned in the previous section, we will not consider weak focusing in the bending magnets. In addition, we neglect effects of the longitudinal fields that result in a focusing strength proportional to the second order of magnetic fields. We consider only effects of the transverse fields. Moreover, we deal only with fields that satisfy the Laplace equation for their potentials (the scalar potential $\varphi$ for the electric fields and the longitudinal component of the vector potential $A$). In other words we assume that

$$\varphi_{xx} + \varphi_{yy} + \varphi_{ss} \approx \varphi_{xx} + \varphi_{yy} = 0, \qquad (4)$$

where the subscript indicates a partial derivative. The same assumption is used for the vector potential as well. It is based on the fact that the third term of the L.H.S. of (4) is (in the lattice construction of the next sections) of the order of $(r_c/\beta)^2 \sim 10^{-4}$, where $r_c$ is the vacuum chamber radius and $\beta$ is the typical beta function value in accelerators. This ratio is comparable to the typical relative field accuracy of accelerator components. In addition, we can always approximate a nonlinear lattice with the smooth dependence of element strengths by constant field sections with a varying strength in a manner similar to difference equation approximations to ordinary differential equations with an arbitrary accuracy. In this approach all the kicks from thin elements obey the 2D Laplace equations and the equations for potentials overall are equivalent to (4).

The transverse motion in our case is uncoupled from the longitudinal one and the longitudinal coordinate $s$ is used as "time" throughout the paper. The corresponding time-dependent Hamiltonian of the transverse motion is

$$H = \frac{p_x^2}{2} + \frac{p_y^2}{2} + U(x, y, s), \qquad (5)$$

where $U(x,y,s)$ is the particle's potential energy, which satisfies Eq. (4). The particle is assumed to have a unit mass and charge.

### III. Special Time-dependence of Fields

References [4] present examples of time, coordinates, and momenta transformation for the Hamiltonian of the type (5) to be transformed to a similar Hamiltonian but with a different potential energy. To demonstrate such a transformation let us assume that we have equal linear focusing in the horizontal and vertical planes and some additional time-dependent potential. The Hamiltonian (5) has the form

$$H = \frac{p_x^2}{2} + \frac{p_y^2}{2} + K(s)\left(\frac{x^2}{2} + \frac{y^2}{2}\right) + V(x, y, s), \tag{6}$$

where $K(s)$ is the linear focusing coefficient. Now we can make a normalized variables substitution, following Eq. (2) with an equation for $\beta$ being

$$\left(\sqrt{\beta}\right)'' + K(s)\sqrt{\beta} = \frac{1}{\sqrt{\beta^3}}, \tag{7}$$

where the differentiation has to be taken with respect to $s$. After we introduce the new "time" $\psi$, which is the betatron phase

$$\psi' = \frac{1}{\beta(s)}, \tag{8}$$

we obtain the new Hamiltonian $H_N$ expressed in new variables

$$H_N = \frac{p_{xN}^2 + p_{yN}^2}{2} + \frac{x_N^2 + y_N^2}{2} + \beta(\psi)V\left(x_N\sqrt{\beta(\psi)}, y_N\sqrt{\beta(\psi)}, s(\psi)\right). \tag{9}$$

Below are the three main ideas of this paper:

1) If the potential $U(x_N, y_N) = \beta(\psi)V(x_N\sqrt{\beta(\psi)}, y_N\sqrt{\beta(\psi)}, s(\psi))$ in Eq. (9) is time-independent, the system has at least one invariant of the motion, namely the Hamiltonian (9) itself in new variables:

$$H_N = \frac{p_{xN}^2 + p_{yN}^2}{2} + \frac{x_N^2 + y_N^2}{2} + U(x_N, y_N); \tag{10}$$

2) The accelerator lattice with equal horizontal and vertical focusing can be organized in several ways. First, one can employ solenoids to provide axially symmetric focusing. Second, consider an element of lattice periodicity consisting of two parts: (1) a drift space, $L$, with exactly equal horizontal and vertical beta-functions, followed by (2) an optics insert, *T*, which has the transfer matrix of a thin axially symmetric lens (Figure 1).

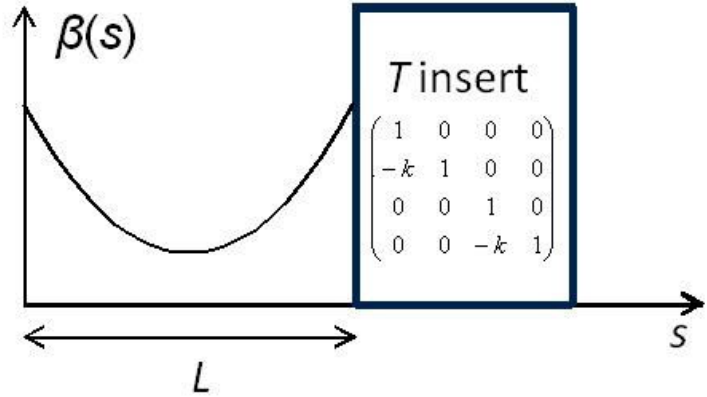

**Figure 1:** An element of periodicity: a drift space with equal beta-functions followed by a *T* insert.

Such an element of periodicity can be organized by regular quadrupoles and dipoles. This lattice structure provides linear focusing. The drift space with equal beta-functions can be used for an additional time-dependent linear or non-linear element characterized by the potential *V(x,y,s)* as in Eq. (6). The linear part (the *T* insert) external to the nonlinear element provides time-independent linear focusing in new variables (9) thus providing stable motion which is otherwise absent in Laplacian time-independent fields;

3) Among all such chosen time-independent Hamiltonians (10) we can find several sets of potentials *U*, which obey the Laplace equation (4) and posses the second integral of motion – such systems are exactly integrable and realizable by magnetic or electric fields!

In the following sections we will give examples of systems with one and two 2D invariants.

### IV. Examples of nonlinear 2D Systems with One Invariant

First, all realizable potentials *U* obey Eq. (4) and can be expressed via one free function:

$$U(x, y) = \operatorname{Re} F(x + iy), \qquad (11)$$

where *F* is any analytical (at least in the vacuum chamber) function of complex variables. The potential *U* in this case automatically satisfies the Laplace equation (4). As a trivial example let us consider an extended quadrupole occupying the entire drift space *L* with the potential $V(x, y, s) = \dfrac{q}{\beta(s)^2}(x^2 - y^2)$, where *q* is the quadrupole magnitude. One can notice that such a potential satisfies Eq. (4) at each *s* location. After the transformation

(2) the new time-independent potential becomes (following Eq. (9)): $U(x_N, y_N) = q(x_N^2 - y_N^2)$. Thus, the motion remains linear and stable (as long as $|q| < ½$).

As a next example, let us consider an extended octupole occupying the entire drift space $L$. It is clear that the octupole strength $V$ has to be inversely proportional to $\beta^3$, $V(x, y, s) = \frac{\kappa}{\beta(s)^3} \left( \frac{x^4}{4} + \frac{y^4}{4} - \frac{3x^2 y^2}{2} \right)$. From (9) we have:

$$U = \kappa \left( \frac{x_N^4}{4} + \frac{y_N^4}{4} - \frac{3 y_N^2 x_N^2}{2} \right), \tag{12}$$

where $\kappa$ is an arbitrary constant. One can see that the resulting potential is time-independent and the Hamiltonian (10) in the normalized variables is the invariant of the system.

Such non-linear potentials, though not integrable, have stable motion around closed orbit, have well defined boundaries, and the escape of particles from the vicinity of a closed orbit is eliminated because the Arnold diffusion is absent in this case (see [5] for more examples and explanations).

### V. 2D Realizable Lattices with Exactly Integrable Motion

Among Hamiltonians (10) there exist potentials that satisfy the Laplace equation and, at the same time, posses the second integral of motion. Here we analyze only systems where this additional integral is quadratic in momenta and, thus, the motion is separable in some variables. This problem has been studied for a long time with the first systematic study carried out by Darboux [6]. We will drop the subscript $N$ for simplicity and will start with the most general time-independent Hamiltonian of the form:

$$H = \frac{1}{2}(p_x^2 + p_y^2) + U(x, y). \tag{13}$$

We will search for the second invariant in the following form:

$$I = A p_x^2 + B p_x p_y + C p_y^2 + D, \tag{14}$$

where $A$, $B$, $C$, and $D$ are functions of $x$ and $y$ only. After coordinate rotations, translations etc. the most general expression for these functions are [7, 8]

$$\begin{aligned} A &= ay^2 + c^2, \\ B &= -2axy, \\ C &= ax^2, \end{aligned} \tag{15}$$

and $D$ is determined from a special partial differential equation along with the potential $U$.

We will now classify the potentials according to values of arbitrary constants, *a* and *c*.

*1. Elliptic coordinates.*

Let $a \neq 0$ and $c \neq 0$, then we will take $a = 1$ and we arrive at the famous Bertrand-Darboux partial differential equation for an integrable potential [6]:

$$xy(U_{xx} - U_{yy}) + (y^2 - x^2 + c^2)U_{xy} + 3yU_x - 3xU_y = 0. \tag{16}$$

This equation has the following general solution:

$$U(x,y) = \frac{f(\xi) + g(\eta)}{\xi^2 - \eta^2}, \tag{17}$$

where *f* and *g* are arbitrary functions and

$$\begin{aligned}\xi &= \frac{\sqrt{(x+c)^2 + y^2} + \sqrt{(x-c)^2 + y^2}}{2c} \\ \eta &= \frac{\sqrt{(x+c)^2 + y^2} - \sqrt{(x-c)^2 + y^2}}{2c}\end{aligned} \tag{18}$$

The second invariant thus yields:

$$I(x,y,p_x,p_y) = (xp_y - yp_x)^2 + c^2 p_x^2 + 2c^2 \frac{f(\xi)\eta^2 + g(\eta)\xi^2}{\xi^2 - \eta^2} \tag{19}$$

First, we would notice that the parabolic potential ($x^2 + y^2$) satisfies Eq. (16) with $f_1(\xi) = c^2\xi^2(\xi^2 - 1)$ and $g_1(\eta) = c^2\eta^2(1 - \eta^2)$. Second, we can solve Eq. (16) together with the Laplace Eq. (4) in elliptical coordinates (see, e.g., [9]) to obtain the following solution:

$$\begin{aligned}f_2(\xi) &= \xi\sqrt{\xi^2 - 1}(d + t\,\text{acosh}(\xi)) \\ g_2(\eta) &= \eta\sqrt{1 - \eta^2}(b + t\,\text{acos}(\eta))\end{aligned}, \tag{20}$$

where *b, c, d* and *t* are arbitrary constants. We omit here the lengthy steps required to obtain Eq. (20). Thus, we arrive at a Hamiltonian exactly in the form of Eq. (10):

$$H(x,y,p_x,p_y) = \frac{p_x^2}{2} + \frac{p_y^2}{2} + \frac{x^2}{2} + \frac{y^2}{2} + \frac{f_2(\xi) + g_2(\eta)}{\xi^2 - \eta^2} \tag{21}$$

in normalized variables (the subscript $N$ is omitted) and with $f_2$ and $g_2$ given by Eq. (20). This system has the second integral in the form of Eq. (19) with $f = \dfrac{f_1}{2} + f_2$ and $g = \dfrac{g_1}{2} + g_2$.

*2. Spherical coordinates.*

If $a \neq 0$ but $c = 0$ we arrive at the following limit of Eq. (21):

$$H = \frac{p_x^2}{2} + \frac{p_y^2}{2} + \frac{r^2}{2} + f(r) + \frac{g(\theta)}{r^2}, \qquad (22)$$

where $\theta$ is the azimuthal angle, and $f$, and $g$ are arbitrary functions at this point. If the potential obeys the Laplace equation, they are:

$$\begin{aligned} f(r) &= d \ln(r) \\ g(\theta) &= b \sin(2\theta) + t \cos(2\theta) \end{aligned} \qquad (23)$$

with $d, b, t$ being arbitrary constants. The second integral in this case is

$$I = (xp_y - yp_x)^2 + 2g(\theta). \qquad (24)$$

*3. Parabolic coordinates.*

The third case (not considered by Darboux) corresponds to $a = 0$. Again, starting from the general expression of Eq. (14) one can show [7, 8] that the second integral has

$$\begin{aligned} A &= 0 \\ B &= -y \\ C &= x \end{aligned} \qquad (25)$$

The equation for the potential $U$ is then

$$2xU_{xy} + 3U_y + y(U_{yy} - U_{xx}) = 0, \qquad (26)$$

which has the solution

$$\begin{aligned} U(x,y) &= \frac{f(r+x) + g(r-x)}{2r}, \\ I &= (yp_x - xp_y)p_y + \frac{(r-x)f(r+x) - (r+x)g(r-x)}{2r} \end{aligned} \qquad (27)$$

where *f* and *g* are arbitrary functions. First, we would notice that the only parabolic solution of Eq. (26) is $(y^2 + 4x^2)$ with $f_1(w) = g_1(w) = w^3$. Thus we can form the axially symmetric and Laplacian focusing terms as

$$\frac{f_1(r+x)+g_1(r-x)}{10r} = \frac{r^2}{2} + \frac{3}{10}(x^2-y^2) = \frac{1}{5}(y^2+4x^2) \tag{28}$$

We will now look for a non-linear potential, which satisfies the Laplace equation as well as Eq. (26). The Hamiltonian has the form

$$H = \frac{p_x^2}{2} + \frac{p_y^2}{2} + \frac{r^2}{2} + \frac{3}{10}(x^2-y^2) + \frac{f_2(r+x)+g_2(r-x)}{2r}, \tag{29}$$

where

$$\begin{aligned} f_2(r+x) &= b\sqrt{r+x} + t(r+x)^2 \\ g_2(r-x) &= d\sqrt{r-x} - t(r-x)^2 \end{aligned} \tag{30}$$

and *b*, *d* and *t* are arbitrary constants. The potential functions *f* and *g* in Eq. (28) are expressed as $f = \frac{f_1}{5} + f_2$ and $g = \frac{g_1}{5} + g_2$.

*4. Cartesian coordinates.*

There are no nontrivial nonlinear solutions separable in Cartesian coordinates. To the authors knowledge, Ref. [3] presents the only known approach to get an uncoupled, nonlinear, regular motion in accelerators in Cartesian coordinates. Its approach differs substantially from this paper.

In addition to the presented above "basic" integrable cases there are systems that can be obtained from the described ones by translations, rotations, etc. We skip these obvious possibilities for brevity. In the next section we briefly analyze their 4-dimensional regions of stability.

## VI. Brief Analysis of Exactly Integrable Cases

All of the above integrable systems are absolutely stable, i.e., all trajectories are bound – none of them escapes to infinity. This follows from the fact that the quadratic term in all normalized variable Hamiltonians is a dominant term at infinity and the motion at large amplitudes is linear and stable. At smaller amplitudes all described above integrable potentials differ substantially from that of presently used multi-pole polynomial potentials – the integrable potentials have singularities at one or two isolated points and their derivatives (forces) may have discontinuities on some intervals or rays. Nevertheless, these potentials satisfy the Laplace equation and can be obtained by appropriately shaped electrodes or magnetic poles. Because of these singularities there is a nontrivial property of these systems - the motion in the transverse plane can be bound in a closed space, or can be bound in a ring-like area, or even more topologically complicated areas. We would like to determine here the dynamic invariants for the case when the motion is bound in the closed area where singularities are absent – this is the simplest case of the vacuum chamber and magnet geometry that is the same as in conventional accelerators (even though the ring-like vacuum chamber shapes with magnets inside is possible as well). All these singularities are simplest in a spherical case when they are just located at *r=0* and are equivalent to those of the singular fields of a line current and a 2D dipole. They are more complex in two other cases. Below we go over three systems from the previous section to state the conditions when the motion is simplest and bound in the closed area without singularities.

*1. Elliptic coordinates.*

We can convert Eq. (21) into Hamilton-Jacobi equation for action *S* and solve for the motion in separated elliptic variables, which is equivalent to replacing momenta in the Hamiltonian (21) with the derivative of the action *S* with respect to the corresponding variable (see, e.g., [10]):

$$p_\xi^2 = \left(\frac{\partial S}{\partial \xi}\right)^2 = 2c^2 E + \frac{\beta - 2c^2 f(\xi)}{\xi^2 - 1},$$
$$p_\eta^2 = \left(\frac{\partial S}{\partial \eta}\right)^2 = 2c^2 E - \frac{\beta + 2c^2 g(\eta)}{1 - \eta^2},$$
(31)

where *E* and *β* are constants (integrals of motion) defined from Eqs. (19) and (21) as

$$E = H(x, y, p_x, p_y)$$
$$\beta = -I(x, y, p_x, p_y) + 2c^2 E$$
(32)

By definition, $|\eta| \leq 1$. In order for the motion to be confined between the two points $\eta = \pm 1$, the energy must be $E \geq 0$ and $\beta + 2c^2 g(\pm 1) > 0$, so that when $\eta$ approaches $\pm 1$, the momentum $p_\eta = \frac{\partial S}{\partial \eta}$ becomes imaginary. These conditions determine the phase space of the motion, bound around zero and not encompassing points $\eta = \pm 1$. For example, take $c = 1$, $f(\xi) = \frac{\xi^4}{2} + 0.25 \xi\sqrt{\xi^2 - 1}$, and $g(\eta) = -\frac{\eta^4}{2}$. This potential has two singularities (poles) at [*x, y*] = [-1, 0] and [1, 0] and its derivative is discontinuous on the interval between these two points. The conditions for the motion to be confined between the poles determine two constraints for the invariants $E \geq 0, \beta > 1$. An example of such a motion is shown in Figure 2. One can see that the trajectory crosses the interval *x* = [-1; 1], *y* = 0. The force is discontinuous on this interval and thus the vertical phase space (the top right figure) has a trajectory derivative discontinuity at *y* = 0. Therefore, such a trajectory may not be physically realized in vacuum.

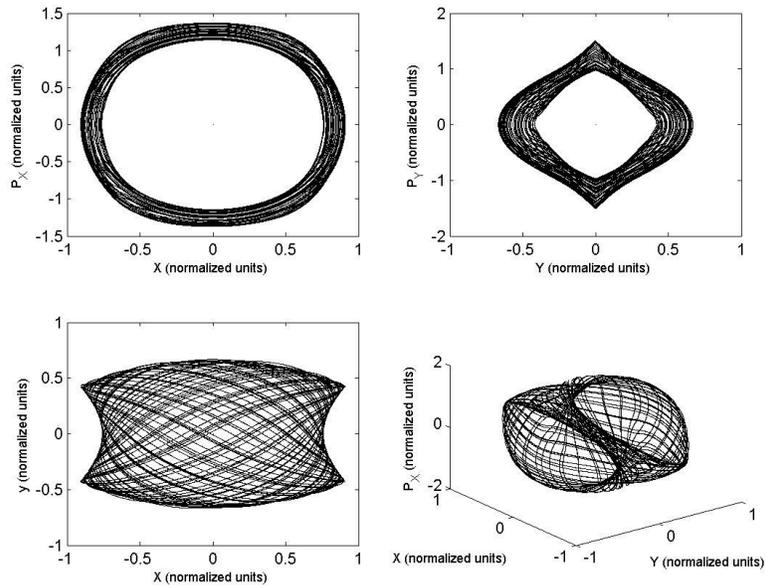

**Figure 2:** Two-dimensional projections of a four-dimensional elliptical-type integrable motion (horizontal phase space (left top), vertical phase space (right top), and x-y projection (left bottom)) and the 3-dimensional trajectory surface in $p_x$, $x$, $y$ coordinates. The initial conditions are $x = 0.9$, $p_x = 0.1$, $y = 0.42$, $p_y = 0.1$, $E \approx 1.91$ and $\beta \approx 1.96$.

Figure 3 shows the opposite case for the same potential. The trajectory never crosses the interval $x = [-1; 1]$, $y = 0$, but the vacuum chamber has to be a ring-like assembly to contain the particles inside.

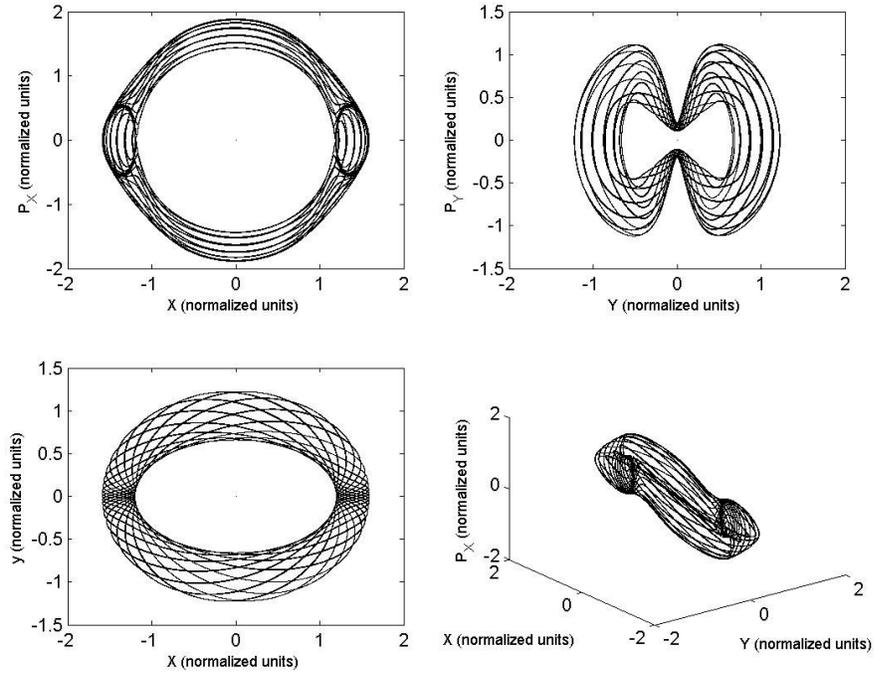

**Figure 3:** Same as Figure 2 except for initial conditions that are taken to be $x = 1.2$, $p_x = 0.1$, $y = 0.001$, $p_y = 0.2$. ($E \approx 3.05$ and $\beta \approx 0.98$).

The Figure 4 presents a different force without a discontinuity between the singularities. It corresponds to the following coefficients in Eq. (20): $c = 1$, $d = 0$, $b = 0.1\pi$ and $t = -0.2$. In the entire volume between singularities the motion is regular, bound, and the fields can be created in a vacuum chamber by electrodes or magnetic poles.

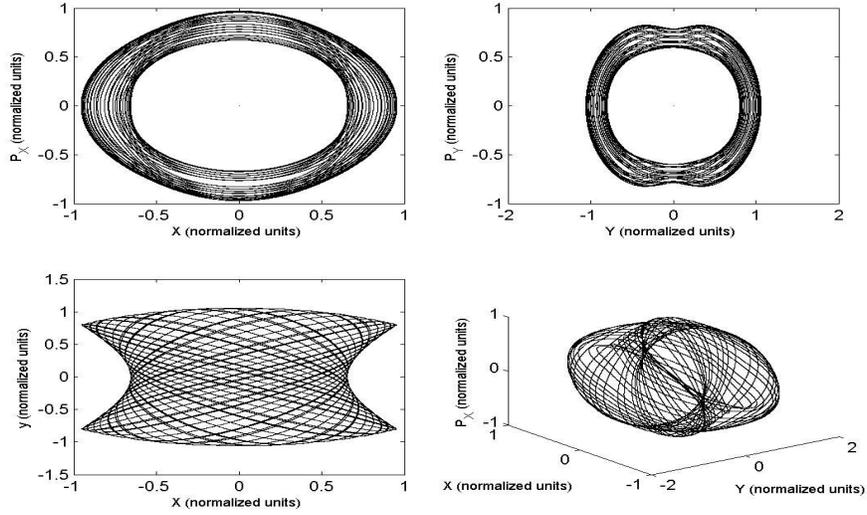

**Figure 4:** Same as Figure 2 except the force coefficients are different and the initial conditions are taken to be $x = 0.95$, $p_x = 0.0$, $y = -0.8$, $p_y = 0.1$ ($E \approx 1.15$ and $\beta \approx 1.356$).

After constructing the action $S$, following Ref. [10] as

$$S(E,\beta) = -E\tau + \int p_\xi d\xi + \int p_\eta d\eta, \qquad (33)$$

where $\tau$ is the time, the motion can be completely solved in quadratures by taking partial derivatives of $S$ with respect to $E$ and $\beta$ and equating these derivatives to new constants of integration.

*2. Spherical coordinates.*

This is the easiest case to analyze. From [10], the solution of the Hamilton-Jacobi equation is (with $p_\varphi=0$ for a 2D case in [10] and functions from Eq. (23)):

$$S = -E\tau + \int \sqrt{\beta - 2b\sin 2\theta - 2t\cos 2\theta}\, d\theta + \int \sqrt{2E - r^2 - 2d\ln r - \beta/r^2}\, dr. \qquad (34)$$

Again, the constants $E$ and $\beta$ are defined from Eqs. (22) and (24) as $E = H$ and $\beta = I$. One can see that the angle $\theta$ can vary from 0 to $2\pi$ if $\beta \geq 2\sqrt{b^2 + t^2}$. This is the condition when the trajectory encircles the singularity $r = 0$; otherwise it covers the area that does not include the singularity. The latter case is presented in Figure 5; the former case is given in Figure 6. The constants in this example are taken: $d = -1$, $b = 0$, and $t = 0.1$. The initial conditions are given in the captions of the Figures 5 and 6.

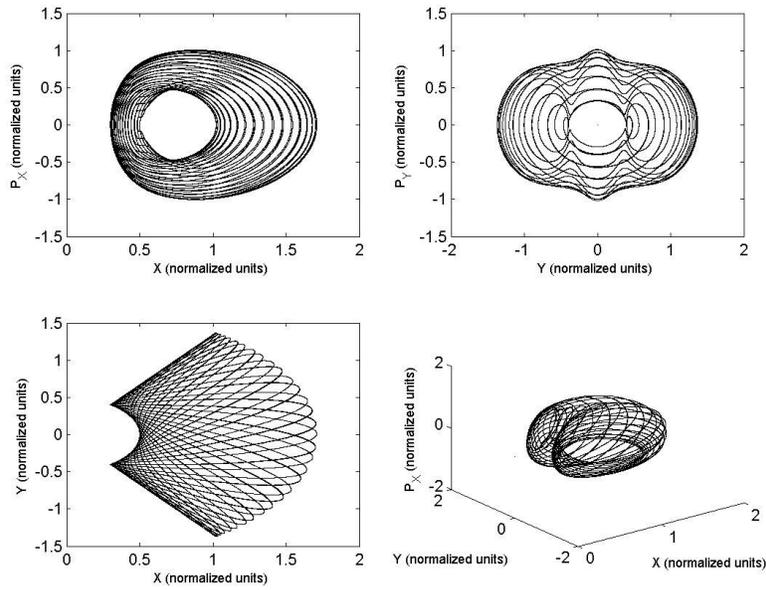

**Figure 5:** An exactly integrable case, separable in spherical coordinates. The four subplots show the same projections as in previous Figures. The initial conditions are $x = 0.3$, $p_x = 0.1$, $y = 0.4$, $p_y = 0.0$, $E \approx -0.675$ and $\beta \approx 0.058$.

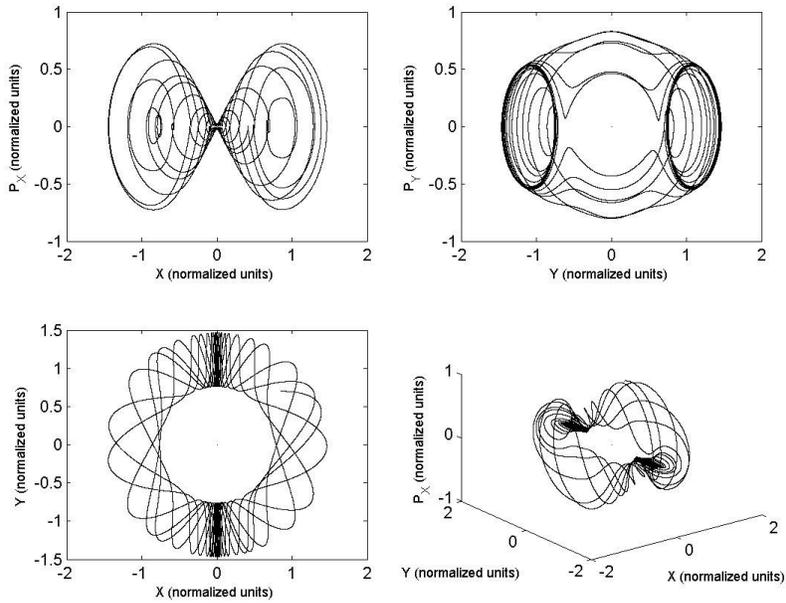

**Figure 6:** An exactly integrable case, separable in spherical coordinates. The four subplots show the same projections as in previous Figures. The initial conditions are $x = 0.85$, $p_x = 0.698$, $y = 0.7$, $p_y = 0.0$. $E \approx 0.962$ and $\beta \approx 0.2004$.

*3. Parabolic coordinates.*

This case is different from the previous two – the trajectories cannot encircle the singularity, $r = 0$. Let us look at the solution of the Hamilton-Jacobi equation from [10] for the functions (30) (again $p_\varphi = 0$ for the 2D case, and $z$ has to be replaced by $x$ in [10]):

$$S = -E\tau + \int \sqrt{\frac{E}{2} + \frac{\beta}{2\xi} - \frac{\xi^2}{10} - \frac{b}{2\sqrt{\xi}} - t\xi}\, d\xi + \int \sqrt{\frac{E}{2} - \frac{\beta}{2\eta} - \frac{\eta^2}{10} - \frac{d}{2\sqrt{\eta}} + t\eta}\, d\eta, \quad (35)$$

where the constants $E$ and $\beta$ are defined from Eqs. (27) and (29) as $E = H$ and $\beta = I$. One can see that for $E > 0$ if $\xi = r+x$ approaches 0 ($\beta > 0$), then $\eta = r-x$ cannot approach zero since the function in the second integral becomes imaginary, and vice versa. It means that the trajectory cannot encircle the singularity. The typical case is shown in Figure 7 for $b = -0.1$, $d = 0$, and $t=0$; the initial conditions are given in the caption. The singularity in this case is at $[x, y] = [0, 0]$ and the discontinuity in the force lies at a horizontal ($x$) negative semi-axis (where $\xi = 0$). The trajectories never cross it if $\beta < 0$ (Figure 7, bottom-left).

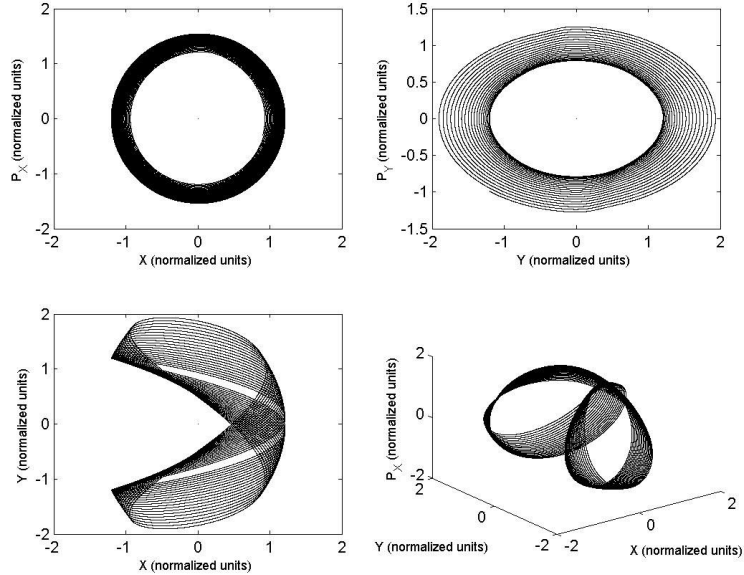

**Figure 7:** Exactly integrable case, separable in parabolic coordinates. The four subplots show the same projections as in previous Figures. The initial conditions are $x = -1.2$, $p_x = -0.1$, $y = 1.2$, $p_y = -0.02$.

In all 3 cases the frequencies in normalized coordinates start from around 1, and then some of frequencies may approach 0 for some initial conditions, which indicates change of topology for the first two cases. Basically, it means that the spread of frequencies in this case can be made around 100% which is important for applications such as, e.g., mitigating instabilities in intense beams.

## VII. Discussion on the Practical Implementation of the Integrable Lattices

As it is clear from the nonlinear lattice construction, the nonlinear portions with the constant Hamiltonian in normalized variables have to be separated by insertions with the transfer matrices $M_x$ and $M_y$, equal for both directions and representing a thin focusing radial kick:

$$M_{x,y} = \begin{bmatrix} 1 & 0 \\ -k & 1 \end{bmatrix}, \tag{36}$$

where $k$ is the arbitrary positive coefficient. Along with the length of the nonlinear section it determines the beta-function behavior (obviously, the constant $k$ and the linear force component of the nonlinear section have to provide stable linear motion). This, in turn, determines the beta-function behavior and gives the dependence of nonlinear fields on longitudinal coordinate. The matrices (36) can be realized as a thin solenoidal lenses or long insertions with few lenses as the space or energy permits. The other possible matrices can have a form:

$$M_{x,y} = \begin{bmatrix} -1 & 0 \\ k & -1 \end{bmatrix}. \tag{37}$$

One can see that the difference of (34) and (35) is that the dynamic variables change sign in the latter case. For motion to be continuous in nonlinear sections, the odd degrees of coordinates in potentials have to change sign after each linear insertion with matrix (37). Overall, the linear insertions have the betatron phase advance equal to the integer of $\pi$, as it was stated in Section III. If the phase advance is zero (solenoidal thin lenses are used as insertions), the spread of betatron frequencies can reach 100% in principle. If it is $\pi$, then it can be made around 50% since the nonlinear section phase advance can approach $\pi$ as well.

The nonlinear section fields can be made by profiling the magnet tips, as described in [3]. The nonlinear element are not required to be strictly continuous but have to approximate the continuous equations with Hamiltonian (10) – the degree of approximation depends on applications and has to be determined at the stage of the nonlinear machine design.

One more note – if one needs to create an intense beam with a very large space charge tune shift it can be done in a self-consistent manner and the system can remain integrable if the space charge force is linear and its gradient is equal in both directions. It can be achieved by making the beam density constant and profiling the vacuum chamber to keep the potential outside compatible with the linear force inside the beam. The paining can be made in the manner suggested in paper [11], that describes various self- consistent 2D and 3D space charge distributions and their realizations in real machines.

## VIII. Conclusion

This paper presents the first finding of exactly integrable nonlinear accelerator lattices realizable with magnetic or electric fields, when the phase space occupied by trajectories has large regions of stability and a very large betatron frequency spread that can mitigate instabilities, space charge effects, and particle loss. In addition, it demonstrates that there exists a variety of nonlinear 2D lattices with one integral of motion which lack Arnold diffusion and can be used for the creation of a bound but chaotic motion or many other integrable cases. Possible lattice constructions are discussed.

**Acknowledgements**

This research is supported by UT-Battelle, LLC and by FRA, LLC for the U. S. Department of Energy under contracts No. DE-AC05-00OR22725 and DE-AC02-07CH11359 respectively.